\documentclass[11pt,a4paper]{article}
\usepackage{amsfonts,amssymb}
\usepackage{cite}
\usepackage[T2A]{fontenc}
\usepackage[cp1251]{inputenc}
\usepackage[english]{babel}
\usepackage{amsmath}%
\usepackage[unicode]{hyperref}
\usepackage{float}

\newtheorem{theorem}{Remark}[section]

\numberwithin{equation}{section}

\title{Integrable Abel equation and  asymptotics
of \\ symmetry solutions of  Korteweg-de Vries equation}
\author{B.I. Suleimanov, A.M. Shavlukov\\
bisul@mail.ru, aza3727@yandex.ru}
\date{}

\begin{document}

\maketitle

\begin{abstract}
We provide a  general solution for a first order  ordinary differential equation   with a rational right-hand side, which arises in constructing asymptotics for large time of
 simultaneous solutions of the Korteweg-de Vries equation and the stationary part of its higher non-autonomous symmetry. This symmetry is determined by a linear combination of the first higher autonomous symmetry of the Korteweg-de Vries equation and of its classical Galileo symmetry. This general solution depends on an arbitrary parameter. By the implicit function theorem, locally it is determined by the first integral explicitly written   in terms of hypergeometric functions. A particular case of the general solution defines self-similar solutions of the Whitham equations, found earlier by G.V. Potemin in 1988.
 In the well-known works by A.V. Gurevich and L.P. Pitaevsky in  early 1970s, it was established that these solutions of the Whitham equations describe the origination in the leading term   of non-damping oscillating waves in a wide range of problems with a small dispersion. The result of this article supports once again  an empirical rule saying that under various passages to the limits, integrable equations can produce only integrable, in certain sense, equations.
 We propose a general conjecture: integrable ordinary differential equations similar to that considered in the present paper should also arise in describing the asymptotics at large times for other symmetry solutions to evolution equations admitting the application of the method of inverse scattering problem.
\end{abstract}

\maketitle

\section{Limiting passages in integrable equations}

According to a well-known empirical law, when describing the asymptotics of solutions to integrable differential equations, as a result of reasonable passage to the limit, there can arise only solutions of integrable
 equations,  but often in another sense  different from the sense of integrability of original equations.

For instance, all possible continuous limits of such integrable differential-difference equations as
Volterra chain 
\begin{align}
(c_n)'_t=c_n(c_{n+1}-c_{n-1})\nonumber
\end{align}
and Toda chain
\begin{align}
(z_n)''_{tt}=e^{z_{n+1}-z_{n}}-e^{z_{n+1}-z_{n}}
\nonumber
\end{align}
can be only differential equations integrable in some sense. The same concerns continuous limits of integrable chain discrete in two independent variables. At present, some of this limiting passages are rigorously justified
\cite{Kal}.

\begin{theorem}
The scientific achievements of R.I. Yamilova made known
first of all owing to his papers 
 devoted to various aspects of the integrability of chains of such kind. In his articles \cite{Yau}-\cite{Gya}, in particular, specific examples of transitions from integrable chains to their continual limits were given.
\end{theorem}

One of the simplest examples of the limiting passage   from one integrable equation to another is the transition from the Korteweg-de Vries (KdV) equation with a small dispersion integrable by the inverse scattering problem method (ISPM)  ($\varepsilon\to 0$)
\begin{align}%\label{SKdV}
u'_t+uu'_x+\varepsilon^2u'''_{xxx}=0,%\label{Skdv}
\nonumber\end{align}
to the Hopf equation
\begin{align} %\label{hopf}
u'_t+uu'_x=0,\nonumber
\end{align}
the general solution of which is locally given by the formula
\begin{align} %\label{shopf}
x-tu=F(u). \nonumber\end{align}
The same Hopf equation is a non-dissipative
limit of the Burgers equation as
 $\varepsilon \to 0$:
\begin{align}u'_t+uu'_x=\varepsilon^2u''_{xx}. \nonumber\end{align}
This equation is reduced to the heat equation $\Lambda'_t=\varepsilon^2\Lambda''_{xx}$ by a linearizing Cole-Hopf change
$u=-2\varepsilon^2 \Lambda'_x/\Lambda$.

A quasiclassical approximation of solutions to an integrable by ISPM nonlinear Schr\"odinger equation (NSE) with a small dispersion
\begin{align}
\label{SKNsh} -i\varepsilon q'_t=\varepsilon^2q''_{xx}+2\delta|q|^2q=0 \qquad (\varepsilon\ll 1)
\end{align}
provides a more complicated example of such passage: 
the substitution
\begin{align}
q=\sqrt{h}\exp{\left(i\frac{\varphi}{\varepsilon}\right)}\nonumber
\end{align}
reduces (\ref{SKNsh}) to a system of two real evolution equations
\begin{align}
h'_{t}+ 2(h \varphi'_x)'_ x= 0,\qquad
\varphi'_{t} + (\varphi'_x)^2 - 2\delta h=\varepsilon^2\frac{( \sqrt{h })''_{xx}}{\sqrt {h} },\nonumber
\end{align}
and their non-dispersion limit is the system
\begin{equation}
h '_{t}+ 2(h \varphi'_x)'_ x= 0,\qquad
\varphi'_{t} + (\varphi'_x)^2 - 2\delta h=0.\label{bsmngo}
\end{equation}
After differentiating the second equation in system
(\ref{bsmngo}) with respect to the variable $x$ and the change $v=2\varphi'_x$, this non-dispersion limit becomes a classical hydrodynamical system
\begin{align*}
h'_t+(hv)'_x=0,\quad v'_t+vv'_x-4\delta h'_x=0;
\end{align*}
the sign of the constant 
 $\delta$ defines its hyperbolic and elliptic version.
Under the hodograph transformation
\begin{align}
v(t,x), h(t,x) \to t(h,v), x(h,v),\nonumber
\end{align}
the solutions of this equation are expressed via the solutions of the linear system of equations
\begin{align}
x'_h=vt'_h+4\delta t'_v, \qquad  x'_v=vt'_v-ht'_h.\nonumber
\end{align}

And a very curious transformation of the meaning of integrability occurs \cite {Zakh}-\cite {Ferkr} when passing to non-dispersion limits of spatially multidimensional integrable ISM equations of the type, for example, the Kadomtsev-Petviashvili equation
\begin{align}\frac{\partial}{\partial t}\left(u'_t+uu'_x+\varepsilon^2u'''_{xxx}\right)
=u''_{yy}.
\nonumber
\end{align}

The integrability of Whitham's hydrodynamic equations by the generalized hodograph method of S.P. Tsarev \cite{Tsa}, \cite{Tsap} \cite{Novdu}, obtained as a result of averaging integrable ISM equations of the KdV, NSE and sine-Gordon type, is also one of the confirmations of the empirical law formulated in the beginning of the paper.
 
For example, using this method in combination with the algebraic-geometric method of I.M.~Krichever~\cite{Kri}, G.V.~Potemin~\cite{Pot}, self-similar solutions of the Whitham equations were found in explicit form, which, according to the well-known results of A.V. Gurevich and L.P. Pitaevskii \cite {g77}, \cite {g78} determine, in the leading order, the behavior as $t\to\infty$ of a universal special solution of the KdV equation
\begin{align} 
\label{KdV} u'_t+uu'_x+u'''_{xxx}=0,
\end{align}
with the asymptotics $u=-x^{\frac{1}{3}}+o(1)$ as $x\pm \infty$.

\begin{theorem}
In \cite{g78}, A.V. Gurevich and L.P. Pitaevsky expressed an opinion that the system of ODEs determining these self-similar solutions of Whitham equations cannot be solved explicitly.
\end{theorem}

Ordinary differential equations (ODEs) of Painlev\'e type integrable by the isomonodromic deformations method \cite{Focn}, often arising in describing the asymptotics at large times for integrable by ISPM equations, are also examples of such limiting equations.

\begin{theorem}
Among the numerous important scientific results of A.B. Shabat, there is also the following one: when describing the asymptotics at large times of a general initial value problem for the KdV equation (\ref {KdV}) in \cite{Sha} and in Section 8.3 of his Habilitation thesis \cite{Shad}, he gave the first example of such transition. 
\end{theorem}

Along with the Painlev\'e equations themselves, in nonlinear problems with a small parameter including those described by solutions of integrable equations, their higher isomonodromic analogs play a universal role in the description of various contrast transition regimes. Both the Painlev\'e equation  and their higher analogs can be regarded as a kind of nonlinear special functions of wave catastrophes.

The general theory of such nonlinear special functions, the foundations of which were laid in the work of A.V. Kitaev \cite{Kit} (simultaneously and independently one particular case, a higher analogue of the second Painlev\'e equation, was considered by the first of the authors of this work in \cite{Lom})  is actively developed today  and it finds numerous applications \cite{Zam}--\cite{Ad}.

In particular, when describing solutions of non-dispersion equations in the vicinity of the folding point, in the leading term   with respect to a small parameter $\varepsilon$ one uses a special universal solution of the Gurevich-Pitaevsky KdV equation (\ref{KdV}), which was discussed in the paragraph above Remark~1.2. And as it was stated in \cite{pjetf}, \cite{s120}, this    special solution at the same time
satisfies
ODE 
\begin{align}u''''_{xxxx}+\frac{5uu''_{xx}}{3}+\frac{5(u'_{x})^2}{6} +\frac{5(u^3-tu+x)}{18}=0,\label{fourp}\end{align}
being the first higher representative of in the hierarchy of isomonodromic analogues of the first Painlev\'e equation in the mentioned paper by A.V.~Kitaev \cite{Kit}.

\begin{theorem}
The same solution of ODE (\ref{fourp}) appears in 
describing of continuous limits of isomonodromic chains
\cite{s120}, \cite{Garsut}. They are related with problems of quantum gravitation theory and were considered in 
   \cite{bresm}, \cite{dous}.
\end{theorem}

In \cite{Kud}, V.R.~Kudashev observed the following fact: after differentiating (\ref{fourp}) with respect to $x$, one obtains a  stationary part of a non-autonomous higher symmetry of the KdV equation (\ref{KdV}), which is determined by the linear combination of the stationary parts of its classical Galilean symmetry $u'_{\tau_G}=1-tu'_x$ and by its first higher autonomous symmetry
 \begin{align} \label{skdp}u'_{\tau_5}=\left(u''''_{xxxx}+\frac{5u''_{xx}u}{3} + \frac{5(u'_x)^2}{6}+\frac{5u^3}{18}\right)'_x.
\end{align}

The symmetry nature of ODE (\ref{fourp}) and the results of paper  \cite{Kudsh} allow one to obtain easily the form of aforementioned solution by G.V.~Potemin of self-similar Whitham equation   \cite{pjetf}, \cite{s120}, \cite{Kud}. In  the last quarter of a century, various properties of solutions to ODE (\ref{fourp}), and mainly, of course, the special solution of Gurevich-Pitaevsky, have been considered from various points of view in many works, see, for example,  \cite{k62}, \cite{Kup}--\cite{GrC}, \cite{Garsut}, \cite{TGl}--\cite{GKC},  
\cite{faa}.

\section{Kudashev integrable equation}

The asymptotics of the universal Gurevich-Pitaevsky solution as   $t\to \infty$ outside the zone of fast oscillation is given by two roots of the cubic fold equation:
\begin{align}\label{Sbor} 
u^3-tu+x=0,
\end{align}
while in the area of non-decaying oscillations it depends on a slow variable   $z=x t^{-\frac{3}{2}}$ and fast phase
\begin{align}\label{Phase} \Phi=t^{\frac{7}{4}}f(z)+f_0(z)\end{align}
and it reads as
\begin{align}\label{Asymp}u=\sqrt{t} (v_0(z,\Phi)+t^{-\frac{7}{4}}v_1(z,\Phi) +t^{-\frac{7}{2}}v_2(z,\Phi)+\dots).
\end{align}
This is almost obviously implied by 
the form
\begin{align}\label{asbor}u=\sqrt{t} v(z) \end{align} of the solution to fold equation (\ref{Sbor}) and the form of linearizations by  (\ref{asbor}) of KDV equation (\ref{KdV}) and ODE~(\ref{fourp}).

In the end of the last century, V.R. Kudashev tried, without employing Whitham averaging method, simply to look for a joint asymptotic solution of   form (\ref{Asymp}) to equations (\ref{KdV}) and (\ref {fourp}). While doing this, he discovered that the function $f(z)$ determining the leading term of the fast phase (\ref{Phase}) by the formula
\begin{align}R(z)=\frac{7f(z)}{4f'_z}-\frac{3z}{2}, \nonumber
\end{align}
is related with a solution of a rather simple first order ODE
\begin{align}
R'_z=\frac{486R^4-171R^2+9zR+5}{9(54R^3-9R+z)(2R+3z)}.
\label{o1}
\end{align}
This equation is obviously equivalent to the Abel equation of second kind:
\begin{align}
(486R^4-171R^2+5+9Rz) z'_R=972 R^4 - 162 R^2 +(1458 R^3- 225 R)z+27 z^2.
\label{ab}
\end{align}

\begin{theorem}
In terms of solutions of equation (\ref{o1}), it is easy to describe the solution of a more general ODE
\begin{align}
r'_x=\frac{486r^4-171tr^2+9xr+5t^2}{9(54r^3-9tr+x)(2tr+3x)}.
\label{ob}
\end{align}
Indeed, as $t=0$, ODE (\ref{ob}) is reduced to ODE $r'_x=r/(3x)$ with the general solution $r=const\cdot x^{\frac{1}{3}}$.  As  $t\neq0$, the solutions of  (\ref{o1}) and (\ref{ob})  are related by the dilatations
\begin{align}r(t,x)=t^{\frac{1}{2}} R(z),\qquad z=x t^{-\frac{3}{2}}.\nonumber\end{align}
\end{theorem}

In \cite{Garsut}, there were written formulae according to which in the case of a special Gurevich - Pitaevsky solution, the solution $R(z)$ of the ODE (\ref{o1}) is related to the self-similar solutions of the Whitham equations from the article by G.V.~Potemin \cite{Pot}. V.R.~Kudashev  knew these formulae but never published them during his lifetime.

 However, after obtaining V.R. Kudashev equations (\ref{ab})  a natural question immediately arose: according to the heuristic law outlined in Section~1 of this article, this equation must undoubtedly be integrable. But how? Below we provide an answer to this question  found last year by using the computer algebra system Maple.
 
Kudashev equation (\ref{ab}) turns out to possess the first integral, which is written explicitly in terms of hypergeometric functions
$_{2}\textrm{\textit{F}}_1(\alpha,\beta;\gamma;y)\equiv F(\alpha,\beta;\gamma;y)$. This integral reads as
($I$ is an arbitrary constant)
\begin{align}
I=\frac{997920 i \sqrt{15} (F(-\frac{7}{6}, -\frac{1}{6};\frac{2}{3};c_1) (a_1 b_6+b_8)+\frac{12789 F(-\frac{1}{6}, \frac{5}{6};\frac{5}{3};c_1) a_1 c_2}{220})}{c_1^{\frac{1}{3}} (F(-\frac{5}{6}, \frac{1}{6};\frac{4}{3};c_1) b_7+77711400 F(\frac{1}{6}, \frac{7}{6};\frac{7}{3};c_1) a_1 (b_5+b_9))},
\nonumber\end{align}
where the functions $a_i$, $b_i$, $c_i$ are given by the formulae 
\begin{align*}
&a_1(R)=R^2-\frac{1}{3},
a_2(R)=R^4-\frac{25}{111}R^2+\frac{1}{1332},
a_3(R,z)=R^3-\frac{2R}{7}+\frac{z}{14},\\
&a_4(R)=R^6-\frac{55}{174}R^4+\frac{19}{1044}R^2-\frac{1}{6264},
a_5(R)=R^6-\frac{503}{3198}R^4+\frac{25}{6396}R^2-\frac{1}{115128},\\
&a_6(R)=R^2(3R^2-1), a_7(R)=-\frac{2665}{1421}R^6+\frac{2515}{8526}R^4-\frac{125}{17052}R^2+\frac{5}{306936},\\
&a_8(R)=\sqrt{3R^4-R^2}, a_9(R)=R^2-\frac{1}{24}, a_{10}(R)=660R^4-76R^2+1,\\
&a_{11}(R)=1980R^6-888R^4+79R^2-1,\\
&b_1(R,z)=2i\sqrt{15} a_8(R) (3R^2-1)+126R^4+9zR-36R^2,\\
&b_2(R)=\sqrt{21}(31968 R^6-8532 R^4 +324 R^2 -1)+a_8(R)(-83160 R^4
+9576 R^2-126),\\
&b_3(R)=-\frac{148 \sqrt{21} a_2(R) a_8(R) a_9(R)}{1155}+R^8-\frac{74}{165}R^6+\frac{79R^4-R^2}{1980},\\
&b_4(R)=-\frac{533 i \sqrt{35} a_1(R) a_5(R)}{1421}+5 a_4(R)R^2, b_5(R)=a_1(R) R^2 \left(3i\sqrt{35}a_5(R)+\frac{4263a_4(R)}{533}\right),\\
&b_6(R)=\frac{1}{220}\left(i \sqrt{35} a_6(R)^{\frac{3}{2}} a_{10}(R)+19188 \sqrt{21} a_3(R) a_5(R) R
-7308 i \sqrt{15}a_1(R) a_4(R) R^2\right),\\
&b_7(R,z)=15120 i \sqrt{15} a_6(R)^{\frac{3}{2}} a_{11}(R)
-124338240 R (i \sqrt{15} a_1(R) R(\sqrt{21}a_1(R) a_5(R)
\\
&\hphantom{b_7(R,z)=}- \frac{1036 a_2(R) a_8(R) a_9(R)}{533})
-\frac{29841a_3(R,z) a_4(R) R^2}{533}+7 \sqrt{21} a_3(R) a_5(R) a_8(R)),\\
&b_8(R,z)=\frac{444 a_8(R)}{385} \left(3i \sqrt{35} a_1(R)^2 a_2(R) a_9(R)-\frac{29841 a_3(R,z) a_4(R) R}{148}\right),\\
&b_9(R)=a_8(R) \left(-\sqrt{21} a_1(R) a_5(R)-\frac{1421 i \sqrt{15} a_4(R) R^2}{533}\right),\\
&c_1(R,z)=-\frac{2993760 i \sqrt{15} a_1(R) b_3(R)}{b_1(R,z) b_2(R)},\\
&c_2(R)=a_8(R) b_4(R)+a_1(R) R^2\left(\sqrt{21} a_7(R)+i \sqrt{15} a_4(R)\right).
\end{align*}

In the following table we provide approximate values of  $I$ corresponding to solutions of five different initial problems for ODE  (\ref{ab}).
\begin{table}[H]
\centering
	\begin{tabular}{|l|l|l|}
		\hline
		Initial value & Approximate value of $I$
		 & Interval of calculation in  $R$ \\ \hline
		\ $z(0)=0$ & \ $0.0960605+0.1663816i$ & \ $R \in [-10, 10]$\\ \hline
		\ $z(1)=0$ & \ $0.0194046+0.0336097i$ & \ $R \in [-11, 11]$ \\ \hline
		\ $z(5)=7$ & \ $0.0308202+0.0533821i$ & \ $R \in [-1, 15]$ \\ \hline
		\ $z(50)=75$ &\  $0.03177193+0.05503056i$ &  \ $R \in [-1, 60]$ \\ \hline
		\ $z(-0.57735)=0.3849$ &\  $0.000591+0.0010237i$ & \ $R \in [-10, 0]$ \\ \hline
	\end{tabular}
	\medskip
\begin{center}
\caption{Approximate values of   $I$ corresponding to five different initial problems.}
\label{t1}
\end{center}
\end{table}

\section{Conclusion}

In the opinion of the authors of the article, its result is potentially not only of a particular importance.

For example, there arises a conjecture that the systems of two non-autonomous ODEs
\begin{align}
\label{urab}
&
\begin{aligned}
&A'_s=\frac{(2A-1)(-288A^3+192A^2+24sA-27A-4s+4B)}{(A+2s)(-576A^3+504A^2-126A+48sA+8B-12s+9)},\\
&B'_s=\left(36A+3s-54A^2-\frac{9}{2}\right)A'_s+\frac{108A^3-108A^2-6sA+6B+27A+4s}{2A+4s},
\end{aligned}
\\
\label{urad}
&
\begin{aligned}
&A'_s=\frac{(2A-1)(288A^3-192A^2+24sA+27A-4s-4B)}{(A-2s)(-576A^3+504A^2-126A-48sA+8B+12s+9)},\\
&B'_s=(36A-3s-54A^2-\frac{9}{2})A'_s-\frac{108A^3-108A^2+6sA+6B+27A-4s}{2A-4s},
\end{aligned}
\end{align}
considered in papers by R.N.~Garifullin \cite{Garri}, \cite{Garut} should also possess  two first integrals, which can be explicitly written in terms of hypergeometric functions.  By means of solutions to systems of ODEs (\ref{urab}) and  (\ref{urad}), one describes the asymptotics as $t\to \infty$ of form
\begin{align}u=t(v_0(z,\Phi)+t^{-\frac{5}{4}}v_1(z,\Phi)+t^{-\frac{5}{2}}v_2(z,\Phi)+\dots), \quad \Phi=t^{\frac{5}{2}}f(s) +n(s) \quad \left(s=\frac{x}{t^2}\right)\nonumber
\end{align}
for joint solutions of KdV equation (\ref{KdV}) and fifth order ODE
\begin{align}\left(u''''_{xxxx}+\frac{5u''_{xx}u}{3} + \frac{5(u'_x)^2}{6}+\frac{5u^3}{18}\right)'_x \pm
\frac{ 2u+xu'_x-3t(u'''_{xxx}+uu'_x)}{6}=0,\nonumber\end{align}
determined by linear combinations of stationary parts of the first higher non-autonomous symmetry  (\ref{skdp}) and a classical dilatation symmetry
$u_{\tau_r}=2u+xu'_x-3t(u'''_{xxx}+uu'_x)$.

We propose a conjecture: a similar asymptotics at large times for joint solutions of various nonlinear integrable by ISPM evolution equations with ODEs of the higher Painlev\'e type  determined by non-autonomous symmetries and, more generally, by invariant manifolds of these evolution equations, should also be described by ODEs integrable in a similar sense.

\end{document}